\documentclass[a4paper,12pt]{article}
\usepackage{graphicx}
\begin{document}
\newcommand{\tchi}{\tilde{\chi}}
\newcommand{\gsim}{\buildrel>\over{_\sim}}
\newcommand{\lsim}{\buildrel<\over{_\sim}}
\newcommand{\psla}{p\kern-.45em/}
\newcommand{\esla}{E\kern-.45em/}
\newcommand{\tl}{\tilde{l}}
\setcounter{page}{0}
\thispagestyle{empty}
\begin{flushright}
UT-ICEPP 99-04 \\
YITP--00--5 \\
January 2000 \\
\end{flushright}

\vspace{2cm}

\begin{center}
{\large\bf Lepton Energy Asymmetry and
 Precision SUSY study at Hadron Colliders}
\end{center}
\baselineskip=32pt

\centerline{
Mihoko M. Nojiri$^a$, Daisuke Toya$^b$ and Tomio Kobayashi$^b$ }

\baselineskip=22pt

\begin{center}
\footnotesize\it

$^a$\, YITP, Kyoto University, Kyoto, 606-8502, Japan \\

$^b$\, ICEPP, Universiy of Tokyo, Hongo, Bunkyo, Tokyo, 113-0033, Japan

\end{center}

\vspace{1cm}

\begin{abstract}
We study the distribution of lepton pairs from the second lightest
neutralino decay $\tchi^0_2\rightarrow \tl l$ followed by
$\tl\rightarrow \tchi^0_1 l$. The distribution of the ratio of lepton
transverse momenta $A_T$ shows peak structure if $m_{ll}\lsim m^{\rm
max}_{ll}/2$ is required. The peak position $A_T^{\rm peak}$ is described by
a simple function of the ino and slepton masses in the $m_{ll}\sim 0$ 
limit. When a moderate $m_{ll}$
cut is applied, $A_T^{\rm peak}$  depends on the
$\tchi^0_2$ velocity distribution, but the dependence would be corrected
by studying the lepton $P_T$
distribution. $A_T^{\rm peak}$
and the edge of $m_{ll}$ distributions are
used to determine the mass parameters involved in the decay
for parameters of interest to LHC experiments.  For some
cases the ino and slepton masses may be determined within 10\% 
by the lepton distribution only independent of  model assumptions. 
Correct combinations of $A_{T}^{\rm peak}$ 
and $m_{ll}^{\rm edge}$ would be identified even if  different
$\tchi^0_2$ decay chains are co-existing. The analysis could be extended to 
the Tevatron energy scale or other cascade decays.
\end{abstract}

\vspace{2cm}

\vfill

\pagebreak

\baselineskip=22pt

\section{Introduction}

The Minimal Supersymmetric Standard Model (MSSM) \cite{SUSY} is one of the
most promising extensions of the Standard Model. It offers a natural
solution of the hierarchy problem, amazing gauge coupling unification,
and dark matter candidates. If Nature chooses  low energy
supersymmetry (SUSY), sparticles will be found {\it for sure}, as
they will be copiously produced at future colliders such as
Large Hadron Collider (LHC) at CERN or TeV scale $e^+e^-$ linear
colliders (LC) proposed by DESY, KEK, and SLAC. LHC would be a great
discovery machine. Squarks and gluinos with mass less than a few TeV
would be found unless the decay patterns are non-canonical
\cite{snowmass}.

On the other hand, the MSSM suffers severe flavor changing neutral 
current (FCNC) constraints if no mass relation is imposed on  sfermion mass
parameters \cite{FCNC}. Various proposals have been made 
for the mechanism to incorporate SUSY 
breaking in ``our sector'', trying to offer natural explanations 
of such mass relations \cite{SUSYB}. 
In short, it would be very surprising
if sparticles are found in  future collider experiments --- The discovery 
is not the final goal, but it is the beginning 
of a new  quest for ``the mechanism'' of SUSY breaking. 

Measurements of soft breaking masses would be an important aspect 
of the study of SUSY, because different SUSY breaking 
mechanisms predict different sparticle mass patterns. 
Studies at the Tevatron and LHC would suffer from substantial
uncertainties and backgrounds compared to an LC, such as luminosity
error, combinatorial backgrounds, and unknown initial energy. While the
discovery of sparticles is guaranteed at the LHC, detailed studies
there would be challenging.  Therefore it is very interesting to see
the ultimate precision of supersymmetric studies at the LHC.

It is possible to determine masses of sparticles 
from the measurement of 
end points of invariant mass distributions \cite{snowmass,HP1,GM,HP2}.
For the minimal 
supergravity  (MSUGRA) and gauge mediated (GM) models, 
there was substantial 
success for the parameter 
points where the decay  of the second lightest neutralino to lepton pair 
$\tchi^0_2\rightarrow ll\tchi^0_1$ is detected with substantial 
statistics. 
For some case, one would be able to not only determine all MSUGRA
parameters, but also to measure the masses of some sparticles,
using the edges and end points of invariant mass distributions involving jets
and leptons. The systematic errors of such analyses may be controlled
if the acceptance near the end points and (jet) energy resolution
are known.
 
Detailed studies in this direction have been performed, and we do not
repeat these here.  In this paper, we instead study the ratio of
lepton $P_T$ (lepton $P_T$ asymmetry $A_T\equiv
P^{l}_{T2}/P^{l}_{T1}$; $P^{l}_{T2}< P^{l}_{T1}$ ) for the decay 
$\tchi^0_2\rightarrow \tl l
\rightarrow ll\tchi^0_1$.  The information has been used in previous
analyses \cite{HP1,IK} in the context of global fits of MSUGRA
parameters.  We show that it is possible to make a direct connection
between the peak structure of the asymmetry $A^{\rm peak}_T$ and the
ratio of the lepton energies in the neutralino rest frame $A_{E}$ by
using events with $m_{ll}<m^{\rm max}_{ll}/2$. We also point out that
systematics due to the $\tchi^0_2$ velocity distribution would be
small and reduced further if one includes the $P_T$ distribution of
the hardest lepton in the fit. Using the $m_{ll}$ end point and the
peak position of the $A_T$ distribution, one can at least determine
two degrees of freedom of the three parameters involved in the
$\tchi^0_2$ decay, $m_{\tchi^0_2}$, $m_{\tchi^0_1}$ and $m_{\tl}$.
The measurements are based on lepton distributions only and free from
uncertainty due to jet energy smearing.

%first 

The organization of this paper is as follows. In section 2, we analyze
the MSUGRA points which were studied in \cite{snowmass,IK}, where
squark and gluino decays are the dominant sources of $\tchi^0_2$.  We
concentrate on the case where $\tchi^0_2\rightarrow\tl l$ is open and
followed by $\tl\rightarrow l\tchi^0_1$.  We find that the $A_T$
distribution has a peak if $m_{ll}\lsim m^{\rm max}_{ll}/2$ is required.  
In the
limit where $m_{ll}\sim 0$, the peak {\it necessarily} agrees with the
ratio of lepton energies $A^0_E = E_{l2}/E_{l1}$ in the $\tchi^0_2$
rest frame for any value of the $\tchi^0_2$ velocity. $A^0_E$ is a
simple function of the ino and slepton masses.  We show that a small
$m_{ll}$ cut promises smaller systematic errors by comparing
distributions for different neutralino velocities.  In section 3, we
show Monte Carlo simulations for the MSUGRA points.  We find nearly
perfect {\it quantitative} agreement between the expectation and MC
data for wide parameter regions. In section 4, we show that
systematics dues to the $\tchi^0_2$ velocity distribution could be
corrected by the hardest lepton's $P_T$ distribution. We also show
expected errors on ino and slepton masses. For the most optimistic
cases where the end point of the lepton invariant mass distribution of
the three body decay $m_{ll}^{\rm 3 body}$ is observed in addition to
the edge of the $m_{ll}$ distribution of the two body decay $m^{\rm 2
body}_{ll}$\llap , we can determine $m_{\tchi^0_2}$, $m_{\tchi^0_1}$ and
$m_{\tl}$ from those (almost) purely kinematical information. At least
two degrees of freedom of the three mass parameters would be determined
by our method if $m^{\rm max}_{ll}\gg 25$ GeV. Section 5 is devoted
to discussions.

\section{ Distribution of lepton energy asymmetry 
with $m_{ll}$ cut}

At hadron colliders, the second lightest neutralino $\tchi^0_2$ 
would be produced in $\tilde{q}$ and $\tilde{g}$ 
decays, or in $\tchi^{\pm}_1\tchi^0_2$ 
pair production. The decay  $\tchi^0_2\rightarrow\tl l$ 
could be a dominant decay mode if it is open. Followed by $\tl\rightarrow
l \tchi^0_1$, the signal consists of a same flavor and opposite
sign lepton pair
associated  with some missing momentum. It is one of the most
promising SUSY signals at hadron colliders. 

The decay process 
$\tchi^0_2\rightarrow \tl^{\pm} l^{\mp}_1$ $\rightarrow \tchi^0_1
l_1^{\pm}l_2^{\mp}$ 
is described by 
two body kinematics and very simple.  The $m_{ll}$ distribution of the 
lepton pair from the $\tchi^0_2$  cascade decay is 
\begin{equation}\label{e1}
\frac{1}{\Gamma}\frac{d\Gamma}{d m^2_{ll}}=  \frac{1}
{(m^{\rm max}_{ll})^2}.
\end{equation}
where
\begin{equation}\label{e2}
m_{ll}^{\rm max}=\frac{\sqrt{(m^2_{\tilde{\chi}^0_2}-m^2_{\tilde{l}})
(m^2_{\tilde{l}}-m^2_{\tilde{\chi}^0_1})}}{m_{\tilde{l}}}\ \ . 
\end{equation}

The decay distribution is flat in $m_{ll}^2$.  The only physical
information we can get from the $m_{ll}$ distribution is therefore the
value of the end point. It constrains one combination of the
three masses involved in $\tchi^0_2$ decay, as one can see in
Eq.(\ref{e2}).

In the rest frame of the second lightest neutralino, the 
energy of $l_1$  is a function of $m_{\tchi^0_2}$ and 
$m_{\tl}$, while $E_{l_2}$ also depends on $m_{ll}$ and $m_{\tchi^0_1}$: 
\begin{equation}\label{e3}
E_{l_1}=\frac{m^2_{\tchi^0_2}-m^2_{\tl}}{2m_{\tchi^0_2}}, \ \ 
E_{l_2}=\frac{m^2_{ll} + m^2_{\tl}-m^2_{\tchi^0_1}}{2 m_{\tchi^0_2}}
\end{equation}
The angle $\theta_{ll}$ between the two leptons in the $\tchi^0_2$ 
rest frame is obtained by solving 
\begin{equation}\label{e4}
m^2_{ll}= 2 E_{l_1} E_{l_2} (1-\cos \theta_{ll})
\end{equation}
$\theta=0$ for $m_{ll}=0$, while $\theta=\pi$ for 
$m_{ll}= m_{ll}^{\rm max}$. 

In Eq.(\ref{e3}), we see that $E_{l_1}$ is monochromatic in the
$\tchi^0_2$ rest frame. As a result, the energies of the two lepton
are, asymmetric.  The ratio of the transverse momenta of the leptons,
which we call the transverse momentum asymmetry $A_T=P^l_{T2}/P^l_{T1}
\ (P^l_{T1}>P^l_{T2})$, provides another information on the decay
kinematics.\footnote{One may also use lepton energy ratio $E_{l1}/E_{l2}$. 
In general, $P_T$ distribution reflects sparticle masses much 
better than energy distribution.}  
However, $P^l_{T1}$ and $P^l_{T2}$ depend on the parent
neutralino momentum, unlike the Lorentz invariant quantity
$m_{ll}$. The $\tchi^0_2$ velocity distribution in turn depends on
$m_{\tilde{q}}$ and $m_{\tilde{g}}$, although the $\tchi^0_2$ decay
distribution in the $\tchi^0_2$ rest frame itself does not depend on them.

This distribution has been used in global fits of MSUGRA parameters;
$A_T$ distribution ``data'' for one MSUGRA point generated by Monte
Carlo simulator are compared to those of different MSUGRA points\cite{HP1, IK}. 
In this model, all sparticle masses depend on a few
universal soft breaking parameters such as $m_{0}$, $M$, $\tan\beta$,
etc.  When we compare different MSUGRA points, we therefore change
both the parameters of the $\tchi^0_2$ decay, $m_{\tchi^0_2}$,
$m_{\tchi^0_1}$ and $m_{\tl}$, and the parameters of $\tchi^0_2$
momentum distributions $m_{\tilde{q}}$ and $m_{\tilde{g}}$ at the same
time. Therefore it was considered to be less important compared to
invariant mass distributions.

However it is possible to make a more direct connection with the first
set of mass parameters $m_{\tchi^0_2}$, $m_{\tchi^0_1}$ and $m_{\tl}$
if a moderate $m_{ll}$ cut is applied \cite{TKN}. When $m_{ll}$ is
small compared to $m_{ll}^{\rm max}$, the lepton and anti-lepton
nearly go in the same direction.  Then the lepton momentum asymmetry
becomes less sensitive to the parent neutralino velocity.  Even after
the smearing due to the boost of $\tchi^0_2$, $A^0_E\equiv
E_1/E_2\vert_{m_{ll}=0}$ still can be extracted from the peak of
$A_T=P^l_{T2}/P^l_{T1}$
\footnote{The lepton from $\tchi^0_2$ decay and $\tl$ decay 
cannot be distinguished in the experiment.  
It is understood that $A^0_E$ means $1/A^0_E$ when $A^0_E$  
exceeds one. };
\begin{equation}\label{e5}
A^{\rm peak}_T ({\rm or\ } 1/A^{\rm peak}_T)  \simeq A^0_E 
\equiv\frac{m^2_{\tchi^0_2}-m^2_{\tl}}{m^2_{\tl}-m^2_{\tchi^0_1}} , 
\end{equation}
therefore $A^{\rm peak}_T$ constrains the mass parameters 
involved in $\tchi^0_2$ decay, just as $m^{\rm max}_{ll}$ does. 
Note $A^0_{E}$ has monotonous dependence on all parameters while the
$m_{ll}$ edge might be accidentally insensitive on $m_{\tl}$.
\footnote{We thank to M. Drees for pointing out this.}

% need more 
\begin{table}[htb]
\begin{center}
\begin{tabular}{|c|c|c|c|c|c|c|c|}\hline
& $m_0$ &  $M_2$ & $m_{\tchi^0_2}$ & $m_{\tl_R}$ & $m_{\tchi^0_1}$ 
&$m^{\rm max}_{ll}$&$A^0_E$
\\ \hline
IK\cite{IK} &100&150&135.5&120.7&65.2 &51.8&0.368\\ \hline
point 5\cite{snowmass,HP1} &100&300&233.0&157.2&121.5&109.1 &0.336 \\ \hline
point 5-2 &115&300&233.2&167.1&121.6 &111.6&0.496\\ \hline
point 5-3 &120&300&233.3&170.6&121.6 &111.6&0.565\\\hline
point 5-4 &125&300&233.3&174.2&121.6 &111.1&0.646\\\hline
\end{tabular}
\end{center}
\caption{\footnotesize Mass parameters and relevant sparticle 
masses in GeV for the points studied in this paper. ISAJET \cite{ISAJET} is 
used to generate sparticle masses. We also show corresponding $m_{ll}^{\rm max}$ 
and $A^0_E$ in the table. } 

\end{table}

In this paper, we study the power of the $A_T$ distribution in the low
$m_{ll}$ region ($m_{ll}< m^{\rm max}_{ll}/2$) to constrain the
kinematics of the cascade decay $\tchi^0_2\rightarrow$ $\tl l
\rightarrow \tchi^0_1 l l $. We chose the points shown in Table 1, but
our method can be applied in generic MSSM studies.  Unlike the common
approach to immediately go into full MC simulations, we first study
the decay distribution for fixed neutralino velocity (labeled by the
boost factor $\gamma_{\tchi^0_2}$ and the pseudo-rapidity
$\eta_{\tchi^0_2}$) $\Gamma ( A_T
(\gamma_{\tchi^0_2},\eta_{\tchi^0_2}))$.
\footnote{The decay distribution $\Gamma$ depends 
on $\gamma$ and $\eta$ through the 
Lorentz boost of all momenta, which is implicitly 
shown as $A_T(\gamma,\eta)$, or  $P^l_T(\gamma, \eta)$.}

The distribution we observe in  experiments 
is expressed  by convoluting the distribution with the 
velocity distribution 
of $\tchi^0_2$, $F(\gamma,\eta)$, as follows;
\begin{equation}\label{e6} 
      d\sigma(A_T)  \equiv   \int d\gamma d\eta \ F(\gamma,\eta)
\Gamma\left(A_T(\gamma,\eta)\right)
\end{equation}
the measured distribution is also affected by cuts on $\esla_T$, 
$M_{\rm eff}$, etc. However it is still useful to know
how $\Gamma(A_T(\gamma_{\tchi^0_2}, \eta_{\tchi^0_2}))$ depends on 
the underlying mass parameters and the $\tchi^0_2$ velocity.

\begin{figure}[htbp]
\begin{center}
\includegraphics[width=7.5cm,angle=90]{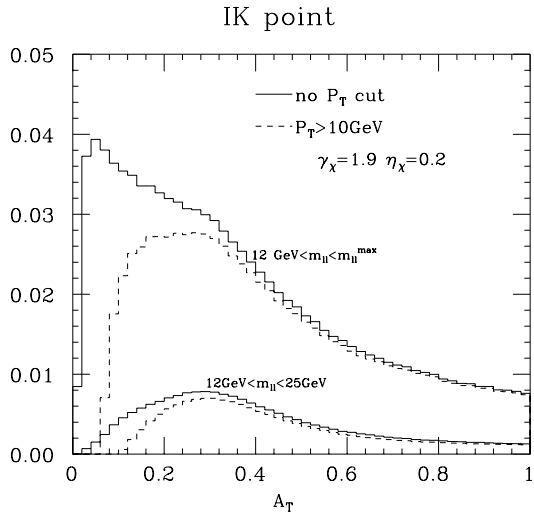}
\end{center}
\caption{\footnotesize  $A_T$ distribution for the $\tchi^0_2$
momentum ($\gamma$,$\eta$)=(1.9, 1.2)  
without $P^l_T$ cut 
(solid), and for $P^l_T>10$ GeV (dashed). The upper histograms are  
without upper $m_{ll}$ cut while the lower histograms are  
distributions with  $m_{ll}<25$ GeV.}
\label{fig1}
\end{figure}

In Fig.~1, we show the $A_T$ distribution with/without invariant mass
cuts and $P^l_T$ cuts. Here we take the IK point and
$\gamma_{\tchi^0_2}=1.9$ , $\eta_{\tchi^0_2}=0.2$.  The distribution
is quite easily obtained by numerical integration.

The distribution without upper $m_{ll}$ cut has some structure 
around $A_T =0.3$ (top solid histogram), but it is insignificant.
(Here we took the events with $m_{ll}>12$ GeV because
large backgrounds from virtual photons are  expected for $m_{ll}<12$ GeV
\cite{BG}). 
With the cut $P^l_T>10$ GeV and the same $\tchi^0_2$ velocity, 
events with  $A_T<0.1$ are hardly accepted, and the distribution 
is roughly flat between  $0.2<A_T<0.3$ (top dashed histogram). 
When the lepton energy in the $\tchi^0_2$ rest frame is small, 
the acceptance efficiency of the events strongly depends on the
velocity  of $\tchi^0_2$, because of the $P^l_T$ cut.
The $A_T$ distribution would  depend on  
the cuts and the distribution of $\tchi^0_2$ velocity
introducing systematical errors to the analysis.

On the other hand, once a moderate $m_{ll}$ cut is applied, the decay
distribution becomes nearly independent of $P^l_{T}$ cuts (bottom
histograms). Here we integrate the region between 12 GeV$<m_{ll}<$ 25
GeV $\sim m^{\rm max}_{ll}/2$.  The distribution has a peak at
$A_T\sim A_E^0=0.368$. The peak is outside the small $A_T$ region
affected by the $P^l_T$ cut. It is also clear from the plot that the
shoulder of the distribution without $m_{ll}$ cut comes from the
events with $m_{ll}<25$ GeV. Note that $\cos\theta_{ll}=0.86  (0.44)$
for $m_{ll}=12 (25)$ GeV in the $\tchi^0_2$ rest frame, therefore
the angle between the lepton and the anti-lepton in the pair is rather
small with the $m_{ll}$ cut.

To put it differently, events above $m^{\rm max}_{ll}/2$
are merely {\it backgrounds} to the $A^0_E$ measurement. 
This is easily understood when we consider the lepton 
configuration near the $m_{ll}$ end point. 
The two leptons go in exactly opposite directions and the asymmetry 
is modified maximally when one of the leptons goes to the direction 
of $\tchi^0_2$ momentum,
$A\equiv E^{\rm lab}_1/E^{\rm lab}_2 =A_E\vert_{m_{ll}=m^{\rm max}_{ll}}
\times \frac{1\pm \beta}{1\mp \beta}$, where $A_E=0.29$ and
$\beta=0.855$ for Fig.~1.
The lepton energy asymmetry in the laboratory  
frame  can range from nearly 0 to 1 due to the boost.

\begin{figure}[htbp]
\begin{center}
\includegraphics[width=7.5cm, angle=90]{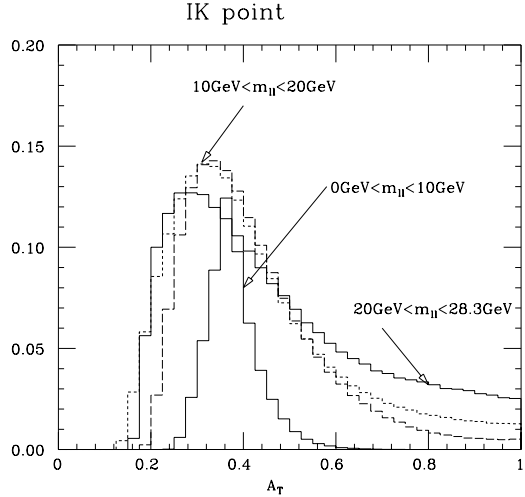}
\end{center}
\caption{\footnotesize $A_T$ distribution under 
different $m_{ll}$ cuts. ($\gamma,\eta$)= (1.4, 0.2), 
and $P^l_{T}>10$ GeV for solid and dotted histogram, while 
the dashed histogram is for $\gamma=2.3$ and $\eta=0.2$.} 
\label{fig2}
\end{figure}
        
It is worth noting that the $A_T$ distribution peaks at smaller value
of $A_T$ as one increases the $m_{ll}$ cut. In Fig.~2, we show
distributions with different $m_{ll}$ cuts, 0 GeV$<m_{ll}<10$ GeV
(solid narrow), 10 GeV $<m_{ll}<20$ GeV (dashed), 20 GeV$<m_{ll}<$28.3
GeV (solid wide), for $\gamma_{\tchi^0_2} =1.4$ and $\eta_{\tchi^0_2}=
0.2$.  The distribution has a sharp peak at a position consistent with
$A_E^0$ for the sample with $m_{ll}<10$ GeV.  $A^{\rm peak}_T=0.323$
for the same neutralino velocity for 10 GeV $<m_{ll}<20$ GeV (dashed
histogram). This shift cannot be explained by  $A_E$ deviation from 
$A^0_E$($A_E=0.363(0.354)$) for $m_{ll}=10(20)$GeV, but it comes from 
the smearing of $A_T$ distribution for finite lepton angle.

The dotted histogram shows a distribution for a higher
neutralino velocity $\gamma=2.3$ and $\eta=0.2$ with 10 GeV
$<m_{ll}<20$ GeV.  The peak position is shifted very little, $A^{\rm
peak}_T\sim 0.321$, therefore it may still be used to determine the
decay kinematics\footnote{Peaks are determined by fitting the
distribution near the peak to a polynomial fitting function.}.  On the
other hand, the distribution off the peak depends more on the
neutralino velocity.  Using the whole distribution introduces a
dependence on the $\tchi^0_2$ momentum distribution, and the fit would
be more assumption dependent.

The distribution is more and more smeared out and peaks at a lower
$A_T$ for larger $m_{ll}$ cuts.  The dependence on the $\tchi^0_2$
momentum is also bigger for the large $m_{ll}$ sample;  $A_T^{\rm
peak} =0.26 \ (0.24)$ for $\gamma=1.4 \ (2.3)$ and 20 GeV
$<m_{ll}<28.3$ GeV. (Only the distribution for the former is shown in
the fugure.)  The distribution is shifted to smaller $A_T$ reducing
the acceptance of the $P^l_T>10$ GeV cut. Some information on the
neutralino velocity distribution is therefore necessary to deduce the
neutralino decay kinematics from the $A_T$ distribution while
increasing the $m_{ll}$ cut in order to increase the statistics and
remove virtual photon backgrounds. This will be discussed in detail in
section 4.

Note that $m_{ll}^{\rm max}\sim 50 $GeV for IK, therefore requiring
$m_{ll}<25$ GeV reduces the number of events in the sample by 1/4. The
reward is a distribution which is less sensitive to $P^l_T$ cuts and
to the $\tchi^0_2$ velocity distribution, and a simple correspondence
to the quantity in the $\tchi^0_2$ rest frame.

\begin{figure}[htbp]
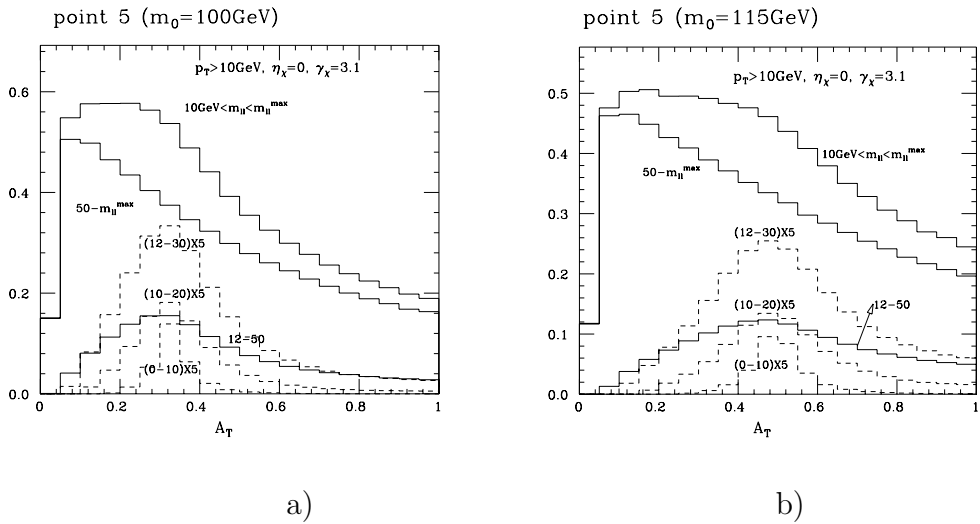

\begin{center}
\hskip -2.5cm 
\includegraphics[width=6.5cm,angle=90]{p5.epsf}
\hskip 0cm 
\includegraphics[width=6.5cm,angle=90]{p5-115.epsf}
\end{center}
\vskip-0.7cm 
\hskip 3.5cm 
a)  
\hskip 6cm 
b)
\caption{\footnotesize $A_T$ distributions for different invariant
mass cuts.  $\gamma_{\tchi^0_2}=3.1$, $\beta_{\tchi^0_2}=0$. 
The distribution for tight $m_{ll}$ cuts are scaled by 
a factor of 5. }
\label{fig3}
\end{figure}

Finally we demonstrate sensitivity of the $A_T$ distribution to the
slepton mass.  We first compare distributions with different slepton
masses, P5 ($m_0=100$ GeV) and P5-2 ($m_0=115$ GeV) in Fig.~3a) and
3b). Here we try a relatively large $\gamma_{\tchi^0_2}$ in order to
have a substantial effect from the $\tchi^0_2$ boost ($\gamma=3.1$
$\eta=0$). Still, the distributions are clearly peaked at $A_T\sim
0.32$ (P5) 0.48 (P5-2) for events with $m_{ll}<50$ GeV, while the
distribution with $m_{ll}> 50$ GeV does not show any structure between
$A_T= 0.1$ to 1. Note $m_{ll}=50$ GeV roughly corresponds to half of
$m^{\rm max}_{ll}$ again.

\begin{figure}[htbp]
\begin{center}
\includegraphics[width=6.5cm,angle=90]{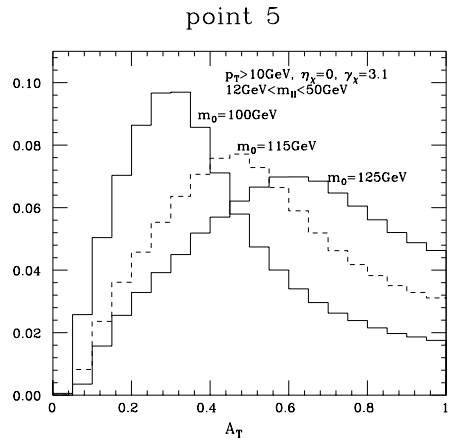}
\end{center}
\label{fig4}
\caption{\footnotesize $A_T$ distributions for different 
slepton masses. Cuts are $12$ GeV $<m_{ll}<50$ GeV, $P^l_T>10$ GeV.  
$(\gamma,\eta)=(3.1,0)$}
\end{figure}

In Fig.~4, we compare distributions with different $m_0$.  Peak
positions shift from 0.3 to 0.63 as one changes $m_0$ by 25 GeV. If
systematic errors are negligible and $M$ is fixed, the sensitivity to
$m_0$ would be $\delta m_0 \sim 1.6$ GeV for $\delta A= 0.02$ (as will
be found in section 3). The peaks are consistent with $A_E^0=0.33$ (for
$m_0=100$GeV), $0.49$ (for $m_0=115$GeV) and 0.65 (for $m_0=125$ GeV).
In section 3, we will find similar agreement for full MC simulation data,
establishing the correspondence.

\section{Monte Carlo simulations}
We are now ready to perform full Monte Carlo simulations to
check the observations made in section 2.

We use ISAJET 7.42\cite{ISAJET} to generate SUSY events.  The
generated events are analyzed by the simple detector simulator
ATLFAST2.21 \cite{ATL}.  The cuts to remove the SM backgrounds down to
a negligible level have already been studied in \cite{HP1,IK}; they
are summarized as,
\vskip 0.5cm
\noindent
{\bf IK (inclusive 3 lepton channel)} \cite{IK};

\noindent
For this point, $M_{\rm eff}$ and $\esla_T$ cuts are not efficient 
because of the light $\tilde g$. A third tagging lepton from $\tchi^0_2$ 
or $\tchi^+_1$ decay is required. When three leptons are in a same flavor, 
the pair of lepton with smaller $\Delta R $ is selected as a lepton pair 
candidate.

\begin{itemize}
\item Two opposite sign same flavor leptons with $P^l_T>15$ GeV. 
\item Third tagging lepton with $P^l_T>15$ GeV.
\item Lepton isolation; No $P_T>2$ GeV track within a 
$\Delta R < 0.3$  cone centered on the lepton track.
\item $\esla_T > 200$ GeV. 
\end{itemize}       	

\noindent
{\bf point 5} \cite{HP1};
\begin{itemize}
\item 4 jets with $P_{T1}> 100$ GeV and $P_{T2,3,4}> 50$ GeV. 
\item $M_{\rm eff}\equiv$ $ P_{T,1} +P_{T,2} +P_{T,3} +P_{T,4} + \esla_T$
$ > 400$ GeV. 
\item $\esla_T> {\rm max}( 100 \ {\rm GeV}, 0.2 M_{\rm eff})$.
\item Two isolated leptons with $P^l_T>10$ GeV, $\vert\eta\vert<2.5$.
Isolation is defined as less than 10 GeV energy deposit 
within a $\Delta R<0.2$ cone centered on the lepton track.
\end{itemize}
We generate $2\times 10^6$ events for each point. This roughly
corresponds to 5 $fb^{-1}$ for IK, and 100 $fb^{-1}$ for point 5. We
present distributions without cuts on $M_{\rm eff}$, jet $P_T$ and
$\esla_T$.  In previous simulations \cite{HP1,HP2,IK}, the acceptance
is roughly constant for all value of $m_{ll}$, therefore those cuts
are expected not to modify the lepton distributions
substantially. Note that substantial acceptance for events with
$m_{ll}<m_{ll}^{\rm max}/2$ is crucial for using the information from
the $A_T$ distribution, as we have seen in section 2.
   
We keep lepton isolation cuts;
\begin{itemize}
\item Less than 10 GeV (15 GeV for IK) 
energy deposit within a  $\Delta R<0.2$ cone centered 
on the lepton track. 
\item No jet within a $\Delta R<0.4$ cone centered on a lepton track. 
\footnote{We use jet finding algorithm of ATLFAST. jet cone size 
is $\Delta R_j<0.4$, Jet finding algorithm requires  
1.5 GeV of minimum energy deposit for the cluster seed, jet cone size 
$\Delta R_j<0.4$, 10 GeV minimum total energy. A resulting  cluster with 
energy more than 15 GeV is called jet.}
\end{itemize}

The acceptance of events turns out to be too high by factor of 3 for
point 5 compared to a full analysis including jet related cuts
\cite{IK,HP1,HP2}. This factor is taken into account when we interpret
the fitting results.\footnote{The number of the selected events for
the IK point is 7000 between 10 GeV to 20 GeV even for the small
luminosity of 5 $fb^{-1}$\cite{IK}. Therefore, the systematic errors
would be dominant for this point.} No plot or fit in this section 
contain SM background, while SUSY background is included.

\begin{figure}[htbp]
\begin{center}
\includegraphics[width=9.5cm]{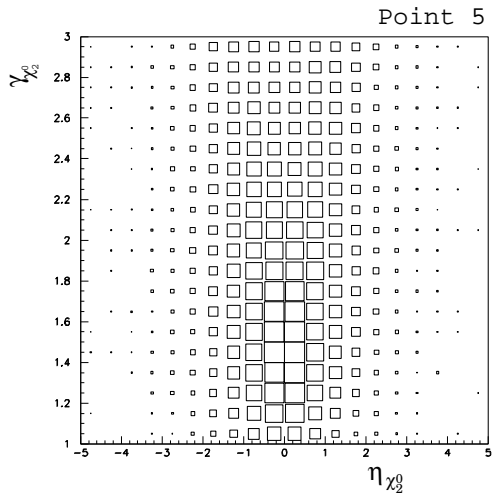}
\end{center}
\label{fig5}
\caption{\footnotesize   
$(\eta,\gamma)$ distribution of $\tchi^0_2$ for point 5.}
\end{figure}

\begin{figure}[htbp]
\begin{center}
\hskip -3.5cm 
\includegraphics[width=9.5cm]{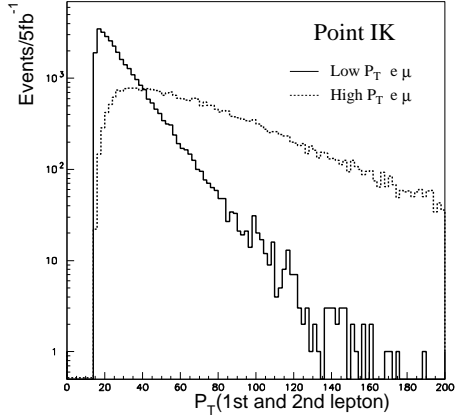}
\hskip -3cm 
\includegraphics[width=9.5cm]{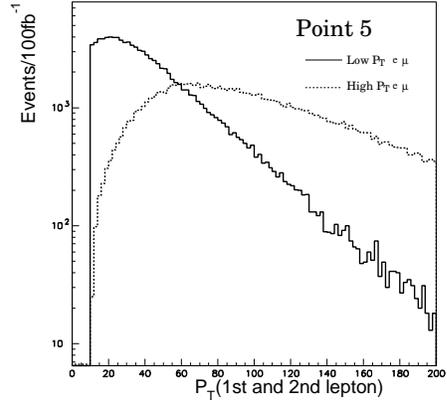}
\end{center}
\vskip-0.3cm 
\hskip 3.5cm 
a)  
\hskip 6cm 
b)
\caption{\footnotesize Lepton $P_T$ distributions of the first (high) and 
the second (low) $P_T$ leptons for a) IK and 
b) pont 5.}
\label{fig6}
\end{figure}

In the previous section, we have already seen that the $A_T$
distribution is somewhat dependent on the parent neutralino velocity
($\gamma_{\tchi^0_2}$, $\eta_{\tchi^0_2}$).  In Fig.~5, we show the
$\gamma_{\tchi^0_2}$ and $\eta_{\tchi^0_2}$ distribution for point IK.
Here one can see that $\eta_{\tchi^0_2}$ is roughly within $\vert
\eta\vert\lsim 1$.  The $\tchi^0_2$ can be very relativistic;
$\gamma_{\tchi^0_2}$ could be much larger than 2.  A modification of
the $A_T$ distribution due to Lorentz boosts is expected unless some
$m_{ll}$ cut is applied. The $P^l_T$ distribution is shown in
Fig.~6. Here we plot the distribution of  higher(lower) of two  lepton $P_T$ 
for dotted(solid) line.  The first(higher) lepton $P^l_T$ can be a few 
times higher than its
most probable value, reflecting the existence of relativistic
$\tchi^0_2$ in the signal sample.

\begin{figure}[htbp]
\begin{center}
\hskip -3.5cm 
\includegraphics[width=9.5cm]{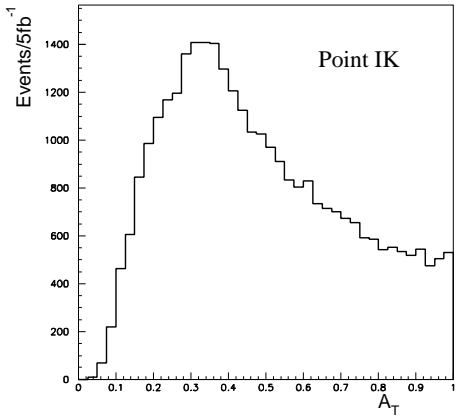}
\hskip -3cm 
\includegraphics[width=9.5cm]{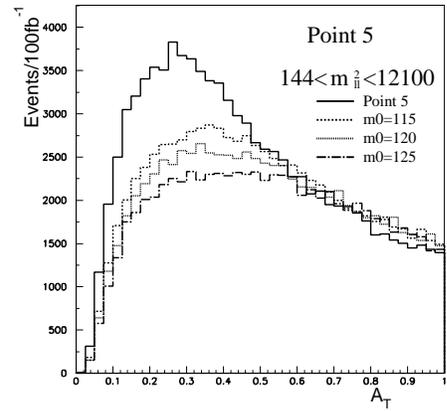}
\end{center}
\vskip-0.3cm 
\hskip 3.5cm 
a)  
\hskip 6cm 
b)
\caption{\footnotesize $A_T$ distribution for a) IK and b) point 5.
without upper $m_{ll}$ cut.  }
\label{fig7}
\end{figure}

We now study the asymmetry distribution in Fig.~7. 
The plot for Point IK (Fig.~7a) shows a smeared peak at 
$A_{T}\sim 0.36$, but the peak is 
rather flat at the top. For point 5 (Fig.~7b), the distribution has 
even less structure, especially when $m_0> 115$ GeV.
Although  global fits of the distributions  must give us information on 
the neutralino decay kinematics, the power to constrain 
neutralino decay parameters would be limited if we try to 
analyze models  without the constraint between soft breaking parameters.

\begin{figure}[htbp]
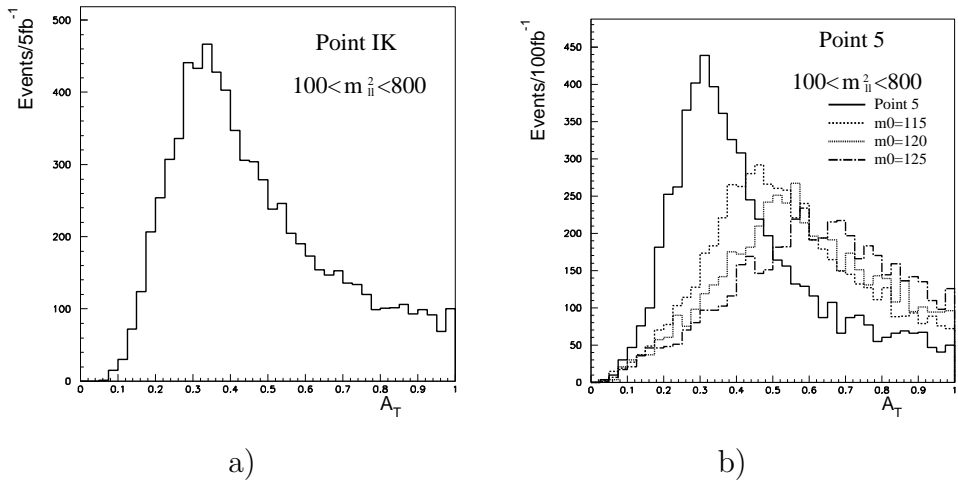

\begin{center}
\hskip -3.5cm 
\includegraphics[width=9.5cm]{ik-mll100-800.epsf}
\hskip -3cm 
\includegraphics[width=9.5cm]{p5-mll100-800.epsf}
\end{center}
\vskip-0.3cm 
\hskip 3.5cm 
a)  
\hskip 6cm 
b)
\caption{\footnotesize  $A_T$ distribution with a
$100 \ {\rm (GeV)}^2<m^2_{ll}<800 \ {\rm (GeV)}^2$ cut }
\label{fig8}
\end{figure}

\begin{figure}[htbp]
\begin{center}
\hskip -4cm
\includegraphics[width=9.5cm]{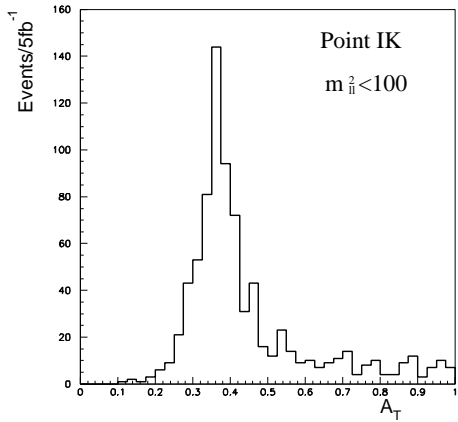}
\end{center}
\label{fig9}
\caption{\footnotesize   $A_T$ distribution for point IK. $m_{ll}<10$ GeV
}
\end{figure}

In Fig.~8, we show $A_T$ distributions with $m_{ll}$ cut, $100 \ {\rm
(GeV)}^2 <m^2_{ll}<800 \ {\rm (GeV)}^2$. We find a narrower peak for
point IK compared to the case without invariant mass cut.  For point
5, improvement of the signal distribution is clear. The peak position
moves right as $m_0$ is increased, and it is consistent with
Fig.~4. Note that for point 5 (IK), $m^2_{ll}<800 \ {\rm (GeV)}^2$
corresponds to $\cos\theta_{ll}=0.72 \ (0.29)$. The angle between the
two leptons is smaller for point 5, which explains the substantial
improvement for point 5.

In Fig.~9, we show the distribution of the events with $m_{ll}<10$ GeV
for point IK. The peak is now nearly delta function like, and it
agrees with $A_E^0$. For point IK, the number of signal events in this
range is statistically significant. If events in this mass range can
be used, we should be able to make a direct $A_E^0$ measurement
without any systematics. However, there could be a significant
background for the events below $m_{ll}<12$ GeV as recently discussed
in \cite{BG}.

We now fit the MC data to  a phenomenological fitting function 
to determine the peak positions and 
the associated errors. The fitting function is chosen as follows:
\begin{eqnarray}\label{e7}
         N(A)&=&N_0 \ 
{\rm exp}\left(-0.5\times\left( \frac{A-A_0}{\sigma}\right)^2\right) 
\ {\rm for}\  A<A_0 \cr
         N(A)&=&N_0\  {\rm exp}\left(-f (A-A_0)\right) 
\ \ \ \ \ \  {\rm for} \ A>A_0
\end{eqnarray}
where  parameters $A_0$, $f$, $N_0$ and $\sigma$  are
determined by minimizing $\chi^2$
using the program MINUIT\footnote{Here we take a completely 
phenomenological assumption for the fitting function, however 
it is much better to use the fitting function based on the 
neutralino velocity distribution calibrated by 
the first lepton $P^l_T$ distribution. See section 4.}.

\begin{figure}[htbp]
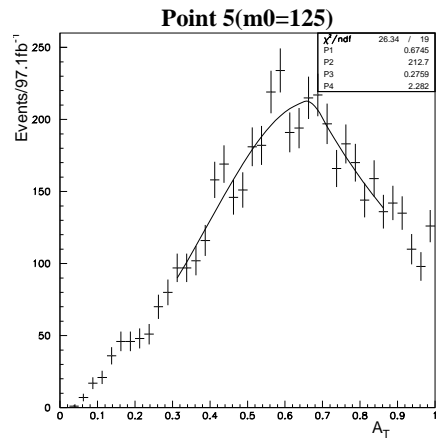

\begin{center}
\hskip -3.5cm 
\includegraphics[width=9.5cm]{fit-ik-mll100-200.epsf}
\hskip -3cm 
\includegraphics[width=9.5cm]{fit-p5-mll100-800.epsf}
\\

\hskip 1.5cm 
a)  
\hskip 6cm 
b)

\vskip 1cm 
\hskip -3.5cm 
\includegraphics[width=9.5cm]{fit-p5-m0115-mll100-800.epsf}
\hskip -3cm 
\includegraphics[width=9.5cm]{fit-p5-m0125-mll100-800.epsf}
\end{center}
\vskip-0.3cm 
\hskip 3.5cm 
c)  
\hskip 6cm 
d)
\caption{\footnotesize Fits to  the MC data using the fitting function 
\ref{e7}. a) IK, b) point 5, c) point 5-2 and d) point 5-4. }
\label{fig10}

\end{figure}

Results of these fits are shown in Fig.~10. For IK (Fig.~10 a)) we fit
the $A_T$ distribution of events with $10$ GeV $< m_{ll}<14.14$ GeV
and find $A_0=0.3408 \pm 0.01$. The peak position is smaller than
$A^0_E=0.36$ (defined in Eq.(\ref{e5})). However, the $A_T$
distribution for fixed neutralino velocity $(\gamma,\eta)= (1.4,0.2)$
indeed peaks at 0.34 if $10$ GeV $< m_{ll}<14.14$ GeV, consistent with
the full MC. As discussed earlier, the peak position does not depend
strongly on the choice of the neutralino velocity when such tight $m_{ll}$
cut is required; this explains the agreement.  For point 5 (Fig.~10, b)-d)),
$A_0= 0.324\pm 0.005$, $0.491\pm0.012$, and $0.675\pm 0.018$ for
$m_0=100,115$ and $125$ GeV, respectively.  The $m_{ll}$ cut
dependence is rather small; $A^0_E$ is $0.33$, $0.49$, $0.65$
respectively, already consistent with the fit\footnote{Note that the
peak positions are at larger $A_T$ compared to Fig.~4. This is because
we select the events below $m_{ll}<28.3$ GeV for Fig.~8, while it is
$m_{ll}<50 $ GeV in Fig.~4 .}. Recall that the total number of events
is a factor 3 too large since we ignored jet related cuts. The
corrected error for 100 $fb^{-1}$ luminosity is 0.009, 0.02 and 0.03
for point 5, 5-2, and 5-4 respectively, assuming statistical scaling.

\section{Model independent mass determination}

The second lightest neutralino might arise from squark and gluino
decays at hadron colliders, therefore the $\tchi^0_2$ velocity
distribution should depend on $m_{\tilde{q}}$ and $m_{\tilde{g}}$.
One may in principle fit the whole distribution to determine model
parameters completely, but various systematic errors could prevent a
complete understanding of the event structure.  We wish to stay with
the distribution which is less model dependent and free of
systematics. Invariant mass distributions are a well established
candidate for such a distribution. In the previous sections we argued
that the peak position of the $A_T$ distribution can be almost
independent of the $\tchi^0_2$ velocity distribution if certain cuts are
applied on $m_{ll}$.

In this section, we will find that the  remaining  minor 
$\gamma_{\tchi^0_2}$, $\eta_{\tchi^0_2}$ distribution dependence may be 
removed by looking into the first lepton $P_T$ distribution.

\begin{figure}[htbp]
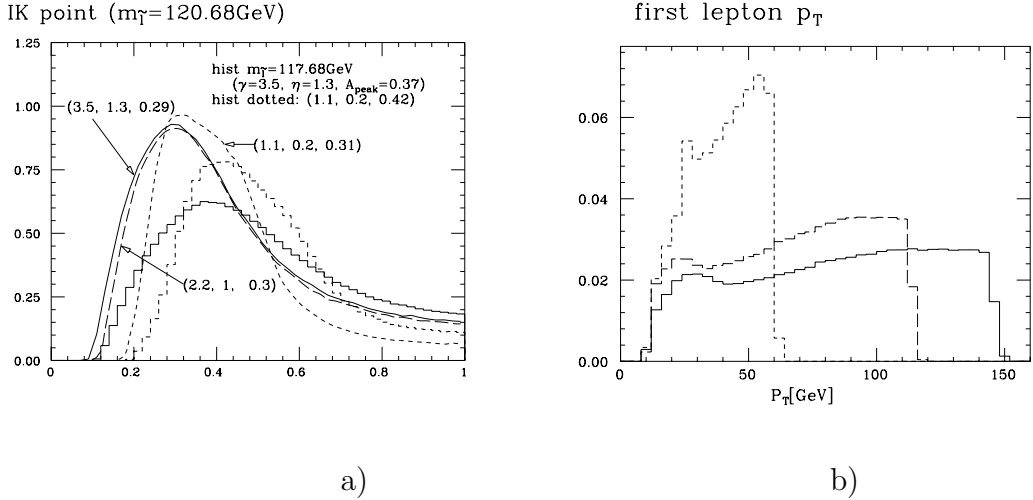

\begin{center}
\hskip -2cm
\includegraphics[width=6cm,angle=90]{betadep-a.epsf}
\hskip 0cm 
\includegraphics[width=6cm,angle=90]{betadep-b.epsf}

\hskip 1.5cm 
a)  
\hskip 6cm 
b)

\caption{\footnotesize a) $A_T$ and b) $P^l_T$ distributions for 
different $\tchi^0_2$ velocity $(\gamma,\eta)$ $=(1.1,0.2), 
(2.2,1)$ and  $(3.5, 1.3)$. $A_T^{\rm peak}$ of each 
distribution is also indicated in the figure.
We also show the histogram for $m_{\tl}=117.68$ GeV in a) for comparison.   }
\label{fig11}
\end{center} 
\end{figure}

In Fig.~11a), we show the $A_T$ distributions for different
$\tchi^0_2$ velocity.  We took point IK and 12 GeV $<m_{ll}<25$ GeV,
therefore the distribution is somewhat dependent on the neutralino
velocity especially when $\gamma_{\tchi^0_2}$ is small. $A_T^{\rm
peak}$ shifts from 0.31 to 0.29 between the representative neutralino
velocity.  In the same figure we also show distributions for
$m_{\tl}=117.68$ GeV. In this case the peak moves from 0.42 to
0.37. The velocity dependence is slightly stronger than for the IK
point.  Although the peak position itself does not depend too much on
the velocity, this certainly suggests some systematics would come into
the fit to the decay parameters.

The $\tchi^0_2$ velocity distribution strongly affects the hardest
lepton $P_T$ distribution, as one can see in Fig.~11b). Here the three
distributions corresponding to Fig.~11a) have totally different
$P^l_{T}$ end points.  We can imagine that the systematics coming from
the neutralino velocity dependence would be substantially reduced if
the $P^l_{T}$ distribution is included in the fit as well.

The $A_T$ and $P^l_T$ distributions can be expressed as convolutions
of the neutralino velocity distribution and neutralino decay distributions
as follows;
\begin{eqnarray}\label{e8}
\sigma(A_T) &=& \int d\gamma_{\tchi^0_2} 
d\eta_{\tchi^0_2} F(\gamma_{\tchi^0_2},\eta_{\tchi^0_2}  )\times 
\Gamma\left(A_T(\gamma_{\tchi^0_2},\eta_{\tchi^0_2} 
) \right),
\\
\sigma(P^l_T) &=& \int d\gamma_{\tchi^0_2}  d\eta_{\tchi^0_2}  
F(\gamma_{\tchi^0_2},\eta_{\tchi^0_2}  )
\times \Gamma\left(P^l_T(\gamma_{\tchi^0_2},\eta_{\tchi^0_2} 
) \right).
\end{eqnarray}
The neutralino velocity distribution $F(\gamma,\eta)$ depends on
parent sparticle masses, while the decay distributions in the
laboratory frame $\Gamma(A_T)$ and $\Gamma(P^l_T)$ depend on
$m_{\tchi^0_1}$, $m_{\tchi^0_2}$ and $m_{\tl}$ implicitly. Various
cuts would be applied to experimental samples of events, therefore
these equations are rather schematic. Note that the two distributions
have different neutralino velocity dependence.  The $\eta_{\tchi^0_2}$
and $\gamma_{\tchi^0_2}$ dependence tend to cancel in $A_T(\gamma,
\eta)$, while the transverse momentum in the laboratory frame
$P^l_T(\gamma,\eta)$ keeps increasing with $\gamma$.  Hence a
measurement of the $P^l_T$ distribution must be very useful for
correcting the minor dependence of the $A_T$ distribution on
$\eta_{\tchi^0_2}$ and $\gamma_{\tchi^0_2}$.

\begin{figure}[htbp]
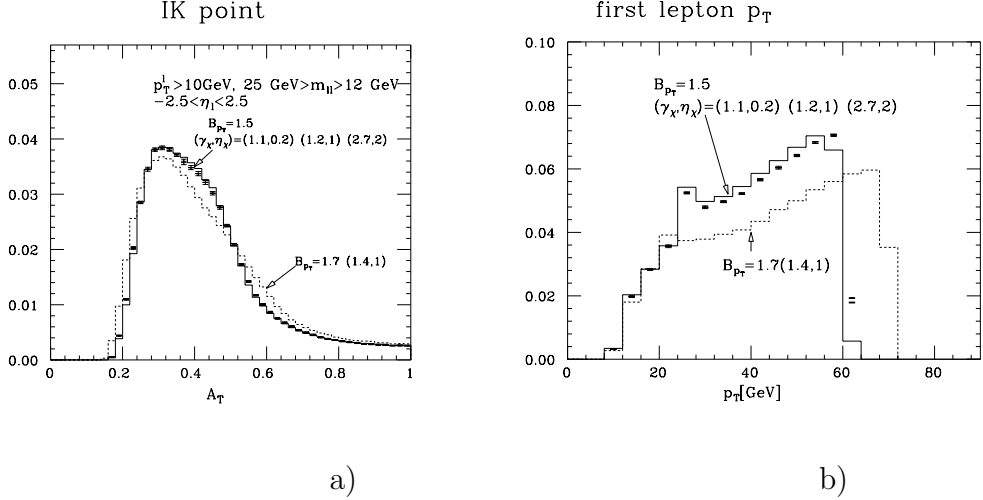

\begin{center}
\hskip -2cm 
\includegraphics[width=6cm,angle=90]{lowbfac-a.epsf}
\hskip 0cm 
\includegraphics[width=6cm,angle=90]{lowbfac-b.epsf}

\hskip 1.5cm 
a)  
\hskip 6cm 
b)

\caption{\footnotesize $A_T$ and $P^l_T$ distributions 
for different neutralino velocities but the same transverse 
boost factor $B_T$. Solid histogram is for $(\gamma,\eta)=(1.1, 0.2)$
while  bar graphs are for $(\gamma,\eta)=(1.2, 1)$ and $(2.7, 2)$ . 
Errors of numerical integrations are shown as bar size. Two 
bars in a same bin are almost coincide.
A distribution 
with a different transverse boost factor is shown by the dotted histogram. }
\label{fig12}
\end{center} 
\end{figure}

The parent neutralino velocity can be decomposed into a boost
$\gamma_T$ from the $\tchi^0_2$ rest frame transverse to the beam
direction, followed by a boost $\gamma_L$ in the beam direction. The
$A_T$ and $P^l_T$ distributions depend on the $\gamma_T$ distribution
while the latter distribution has no effect on them. This can be seen
in Fig.~12 a) and b).  We show three $A_T$ and $P^l_T$
distributions, for $\tchi^0_2$ $(\gamma_{\tchi},\eta_{\tchi})= (1.1,
0.2)$, $(1.2,1)$ and $(2.7, 2)$.  The 3 points have a common feature,
\begin{equation}\label{e10}
B_T(\gamma,\eta)
\equiv \frac{P^{\rm l}_{T}\vert_{\rm max}}
{E_{l1}({\rm at}\  \tchi^0_2 \ {\rm rest})}=
1.5.
\end{equation} 
The $P^l_T$ and $A_T$ distributions of leptons are very similar 
as one can see in Fig.~12 a) and b).

This observation is based on a numerical integration which now takes
into account the cut $\vert \eta_l\vert <2.5$, in addition to $12$ GeV
$<m_{ll}<25$ GeV, and the $P^l_T>10$ GeV cut. The effect of the $\eta_l$
cut turns out to be very small.  We checked numerically that the
distributions with common $P^l_T$ end points are roughly the same with these
cuts. On the other hand, the $A_T$ distribution has significant $B_{T}$
dependence as one can see from the distributions for $B_T=1.7$ (dotted
histograms). This suggests that one only has to know the $\gamma_T$
distribution, which could be reconstructed from the $P^l_T$ distribution.
Schematically, one can write
\begin{eqnarray}\label{e11}
\sigma(A_T) &=& \int dB_T F(B_T)\times \Gamma(A_T(B_T(\eta, \gamma))),
\\
\sigma(P^l_T) &=& \int dB_T F(B_T)\times 
\Gamma(P^l_T(B_T(\eta, \gamma))).
\end{eqnarray}
$\Gamma(A_T(B_T))$, and $\Gamma(P^l_T((B_T))$  are implicit functions 
of ino and slepton masses, and one can fit to experimental data 
to obtain those mass parameters  in addition to $F(B_T)$.
Of course, one must also study the effect of $\esla_T$, $M_{\rm eff}$, 
and $P_{T j}$ cuts and detailed MC simulations are necessary.

Given the indication that the dependence on the $\tchi^0_2$ velocity
distribution can be corrected directly from the $P^l_T$ distribution,
we now use the error on $A^{\rm peak}_T$ and $m_{ll}$ end points to
determine $m_{\tchi^0_2}$, $m_{\tchi^0_1}$, and $m_{\tl}$.  As we have
seen in the previous sections, $A^{\rm peak}_T$ depends on the
$m_{ll}$ cut, but we assume the statistical uncertainty of $A^{\rm
peak}_T$ can be translated into that of $A^0_E$; i.e., we assume that
the correlation caused by only using events within a finite range of
$m_{ll}$ values would be small.

\begin{figure}[htbp]
\begin{center}
\includegraphics[width=7.5cm]{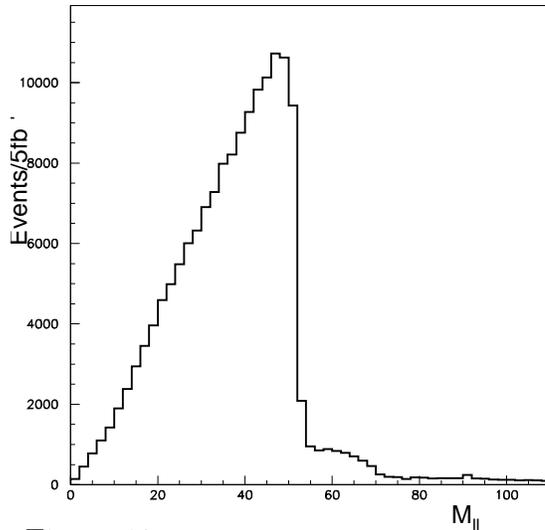}
\end{center}
\label{fig13}
\caption{\footnotesize The $m_{ll}$ distribution 
for point IK.  
}
\end{figure}

We take the IK point as an example; the point is interesting because
both the edge of the $m_{ll}$ distribution due to the two body cascade
decays $m_{ll}^{\rm 2 body}$ and the end point of the three body decay
$\tchi^0_2\rightarrow l l \tchi^0_1$, $m_{ll}^{\rm 3 body}$, can be
seen. (See Fig.~13.)  This is because the right handed slepton coupling to
wino and higgsino is essentially zero, therefore the two body decay coupling is
suppressed.  The measurements of $m_{ll}^{\rm 2 body}$, $A_T^0$, and
$m_{ll}^{\rm 3 body}$ $\equiv$
$m_{\tilde{\chi}^0_2}-m_{\tilde{\chi}^0_1}$ are potentially sufficient
to determine all sparticle masses involved in the $\tilde{\chi}^0_2$
cascade decay.

\begin{figure}[htbp]
\begin{center}
\includegraphics[width=7.5cm, angle = -90]{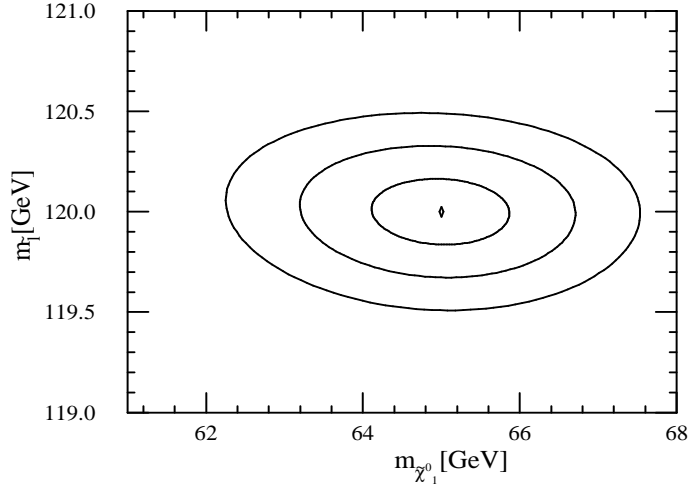}
\end{center}
\label{fig14}
\caption{\footnotesize Contours of constant $\Delta \chi^2=1,4,9$ for
the IK point. We set $\delta m_{\tchi^0_2}=0 $, $\delta m_{ll}^{\rm 2
body}$ 0.5 GeV and $\delta A^0_E= 0.007$. Only the contours 
near the input value are shown. }
\end{figure}

In order to demonstrate the importance of the measurement of $A_T$, we
first show the expected constraints on $m_{\tilde{l}}$ and
$m_{\tilde{\chi}^0_1}$ when $m_{\tilde{\chi}^0_2}$ is fixed. We assume
that $A_T$ and $m_{ll}^{\rm 2body}$ are measured within errors of 0.007
and 0.5 GeV, respectively (Fig.~14). $\Delta\chi^2$ is defined as
\begin{equation}\label{dm}
\Delta\chi^2=\left(\frac{A^0_E-A^{0'}_E}{\delta A^0_E} \right)^2 + 
\left( \frac{m_{ll}-m^{'}_{ll}}{\delta m^{\rm 2body}}\right)^2.
\end{equation}
Here, $A^0_E$, $\delta A^0_E$ and $m_{ll}\equiv m^{\rm 2body}_{ll}$
and $\delta m^{\rm 2body}_{ll}$ are 'data' and the error, while,
$A^{0'}_E$ and $m^{'}_{ll}$ are functions of the ino and slepton
masses.  The resulting $\Delta\chi^2=1,4,9, \dots$ contours roughly
correspond to $1\sigma, 2\sigma, 3\sigma, \dots$ errors on the
parameters.  The errors on $m_{\tilde{l}}$ and $m_{\tilde{\chi}^0_1}$
could be of the order of 1\% or less, consistent with the previous fits in
\cite{IK}. Note, however, that they did not identify the origin of the
peak structure and used the {\it whole} $A_T$ distribution for the
fit.  As we have stressed, this fit will depend on assumptions about
parent squark and gluino masses, while our fit relies solely on the
peak position, directly constraining $m_{\tilde{\chi}^0_1}$,
$m_{\tilde{\chi}^0_2}$ and $m_{\tilde{l}}$.

For the IK point, one can also determine $m^{\rm 3body}_{ll}$.  The
errors on the masses under the three constraints would be
substantially larger than those shown in Fig.~14 (where $m_{\tchi^0_2}$
is fixed).  This is due to correlations between the constraints. This
can be seen in Fig.~15, where $m_{\tchi^0_2}$ and $m_{\tl}$ are shown as
functions of $m_{ll}^{\rm 3 body}$ for fixed values of $A^0_E$ and
$m_{ll}^{\rm 2body}$. Even in the limit where $A^0_E$ and $m_{ll}^{\rm
2 body}$ are known exactly, an error on $m^{\rm 3 body}_{ll}$ of the
order of 1 GeV would result 5 GeV errors on $m_{\tchi^0_2}$ and
$m_{\tl}$.

\begin{figure}[htbp]
\begin{center}
\includegraphics[width=7.5cm,angle=90]{endpoint.epsf}
\end{center}
\label{fig15}
\caption{\footnotesize $m_{\tchi^0_2}$ and $m_{\tl}$ 
as the function of $m^{\rm 3 body}_{ll}$ when 
$A^0_E=0.37$ and  $m_{ll}^{\rm 2 body}=51.9$ GeV.
}
\end{figure}

Assuming an error on $m_{ll}^{\rm 3\ body}$, $\delta m_{ll}^{\rm 3\
body}= 1$ GeV,\footnote{Our assumptions of the errors for 
$m_{ll}$ endpoints are substantially 
conservative to those found in literature\cite{HP1,HP2}}
 $m_{\tilde{\chi}^0_1}$, $m_{\tilde{\chi}^0_2}$, and
$m_{\tilde{l}}$ are constrained within $\sim \pm 8$ GeV, without
assuming any relation between ino and slepton masses. The error is
large compared to those expected from LC experiments, however it still
makes an impressive case where sparticle masses are determined without
relying on model assumptions. \footnote{For point 5, end points of 
$m_{ll}$, $m_{lq}$, $m_{llq}$ distributions in addition to the lower
end point of $m_{llq}$ distribution when  $m_{ll}>m_{ll}^{\rm max}/2$ is 
required determine $m_{\tchi^0_1}$ mass within $O(10\%)$ model 
independently.\cite{HP2}}

Note that the $m_{\tilde e}/m_{\tilde \mu}$
ratio would be constrained strongly.  Assuming $\delta A_T<0.007$,
$\delta m_{ee, \mu\mu}<0.5$ GeV, $\delta m_{ll}^{\rm 3 body}= 4$ GeV,
we obtain $\delta(m_{\tilde e}/m_{\tilde \mu})=2.5 $ \% for $\Delta
\chi^2 <1$, and 7\% for $\Delta\chi^2<9$.
  
Several comments are in order. The background in the region $m_{ll}\ll
m_{ll}^{\rm max}$ must be studied carefully. For example, SM
$t\bar{t}l\bar{l}$ production could be important in the low $m_{ll}$
region. Note that full amplitude level studies of $W\gamma^*$
production have been performed for the background process of
$\tilde{\chi}^0_2$ $\tilde{\chi}^{\pm}_1\rightarrow 3l$, and large
background was found in the $m_{ll}<10$ GeV region \cite{BG}. It has
also been pointed out that $\Upsilon$ production is an important
source of background when $m_{ll}<12$ GeV. However it is unlikely that
the background distribution has a peak at $A_T \gg 0$. A peak of
the signal distribution may be observed precisely on the top of such
backgrounds, especially when signal rates are high enough to allow
precision studies. Besides, one only needs to require
$m_{ll}<m_{ll}^{\rm max}/2$ to see structure in the $A_{T}$ distribution.
The peak position that may deviate from $A^0_E$ could be corrected
from the $P^l_T$ distribution in an almost model independent way.

Recently, it was pointed out in \cite{HP2} that one can obtain the
same information by taking the ratio of the end points of the
invariant masses of jet and lepton(s).  Their analysis was carried out
for point 5. The dominant cascade decay process is
$\tilde{q}\rightarrow \tilde{\chi}^0_2q$ followed by
$\tilde{\chi}^0_2\rightarrow\tilde{l}l_1$, and
$\tilde{l}\rightarrow\tchi^0_1 l_2$.  Jets from squark decays are
substantially harder than the other jets, and can be identified. A
correct set of a jet and a lepton pair originating from a squark decay
is then selected by requiring that $m_{llj}<600$ GeV for one of the two
hardest jets $j$, and $m_{llj'}>600$ GeV for the other jet $j'$.  The
end points of the invariant mass distribution $m_{l_1q}$ and
$m_{l_1l_2q}$ are expressed as simple analytical functions of
$m_{\tilde{q}}$, $m_{\tilde{l}}$, $m_{\tchi^0_{1(2)}}$. One can
reconstruct the $m_{l_1q}$ end point by choosing the combination of
the first lepton and the jet.\footnote{
Note that the efficient selection 
of the first lepton for $m_{lq}$ distribution relies on large lepton
energy asymmetry. However as $A\rightarrow 1$, the end points of 
$m_{l_1 j}$ and $m_{l_2 j}$ tend to coincide, therefor it may not be 
a problem.}

Although  each end point is 10\% to 4\%  
smaller than expectation depending on jet definition, the 
ratio
\begin{equation}
\frac{m^{\rm max}_{lq}}{m^{\rm max}_{llq}}
=\sqrt{\frac{m^2_{\tchi^0_2}-m^2_{\tl}}
{m^2_{\tchi^0_2}-m^2_{\tchi^0_1}}}= \sqrt{\frac{1}{1+(A^0_E)^{-1}}}
\end{equation}
agrees with the expectation.\footnote{ When
$m^2_{\tilde{l}}-m^2_{\tchi^0_1}>$ $m^2_{\tchi^0_2}-m^2_{\tl}$,
$m_{lq}/m_{llq}= \sqrt{(m^2_{\tl} - m^2_{\tchi^0_1}) /
(m^2_{\tchi^0_2} - m^2_{\tchi^0_1}) }$.}  The fitted value ranges from
0.87 ($\Delta R=0.4$ for jet definition) to 0.877 ($\Delta R=0.7$)
while the expectation is 0.868. The range corresponds to $A^0_E=
0.321$ to 0.30 while the expectation is 0.327. Our fit gives
$A_0=0.324\pm 0.009$ for the same point. The comparison of systematics
might be an interesting topic for future studies. Our $A_T^{\rm peak}$
analysis may be performed even if jets and leptons in the same cascade
decay chain can not be identified, therefore it can be applied in a
wider context.
\footnote{Another potential problem of the analysis in
\cite{HP2} is that $m_{\tilde{q}}-m^{\rm max}_{llj}= 145$ GeV is
almost as small as $m_{\tchi^0_1}=122$ GeV. Being the end point of the
$m_{llj}$ distribution requires the $\tchi^0_1$ from the decay chain
to be very non-relativistic in the $\tilde{q}$ rest frame. This should
reduce $\esla_T$ toward the end point. In general the $m_{llj}$ and
$m_{lj}$ end points correspond to different kinematical
configurations; attention must be paid to the consequence for relative
efficiencies.}

It is also interesting to reconstruct the kinematics when both
$\tchi^0_2\rightarrow \tl_R l $ and $\tchi^0_2\rightarrow \tl_L l$ are
open and the branching ratios are of the same order. In addition to
the two edges of the $m_{ll}$ distribution $m^{\rm max}_{ll}({\rm
low})$ and $m^{\rm max}_{ll}({\rm high})$, one should be able to
observe two peaks in the $A_T$ distribution $A_T^{(1)}$ and
$A_T^{(2)}$, corresponding to the two decay chains.  By comparing
$A_T$ distributions for $m_{ll}\lsim m_{ll}({\rm low})/2$ and
$m_{ll}({\rm low})/2<m_{ll}<m_{ll}({\rm high})/2$, one should be able
to determine proper sets of the $m_{ll}$ edge and the peak,
because the peak at $A^{(1)}_T$ can be hardly observed for
$m_{ll}>m_{ll}({\rm low})/2$, while the peak at $A^{(2)}_T$ can still
be seen. Note that there are four parameters for four constraints in this
case, therefore one can in principle solve for all mass parameters.

\section{Discussion}

The second lightest neutralino $\tchi^0_2$ would be copiously produced
from $\tilde{q}$ and $\tilde{g}$ decays at the LHC, and $\tchi^+_1
\tchi^0_2$ production is an important mode for the Tevatron. In this
paper, we have studied the distribution of the $P^l_T$ asymmetry,
$A_T\equiv P^l_{T2}/P^l_{T1}$, of the lepton-anti-lepton pair that arises
from the cascade decay $\tchi^0_2\rightarrow \tl l \rightarrow
\tchi^0_1 ll$.  We have found that the $A_T$ distribution shows a clear peak
structure in a wide parameter region if $m_{ll}<m^{\rm max}_{ll}/2$
is required. The peak position is insensitive to the parent
$\tchi^0_2$ velocity distribution, and in the limit of $m_{ll}\sim 0$,
it is understood as $A_E^0$, the ratio of lepton and anti-lepton
energy in the rest frame of $\tchi^0_2$. The ratio $A_E^0$ is a simple
function of $m_{\tchi^0_2}$, $m_{\tl}$ and $m_{\tchi^0_1}$.

We have also performed MC simulations for several representative
points.  Values of the peak position obtained by fitting MC data agree
with those for $\tchi^0_2$ with typical velocity. This follows from
the insensitivity of the $A_T$ distribution to the parent neutralino
velocity. The typical velocity could be estimated easily by using the
hardest lepton $P_T$ distribution. Therefore the $A_T$ peak can be
used to constrain $m_{\tchi^0_2}$, $m_{\tl}$ and $m_{\tchi^0_1}$.  By
using the edge of the $m_{ll}$ distribution in addition to the $A_T$
distribution, one can determine two degrees of freedom of the three
mass parameters involved in the $\tchi^0_2$ cascade decay. When the
end point of $m_{ll}$ distribution of the three body decay
$\tchi^0_2\rightarrow\tchi^0_1 ll$ can be measured simultaneously, one
can determine the {\em all} mass parameters describing $\tchi^0_2$
cascade decays. The analysis is entirely based on lepton distributions and
does not rely on jet energy measurements.

The reconstruction of the $\tchi^0_2$ momentum distribution is of some
importance for our analysis. The hardest lepton $P_T$ distribution
should allow us to study the $\tchi^0_2$ velocity distribution
independently from the $\tilde{q},\tilde{g}$ mass determination. In
fact, the measurement of this distribution may allow one to constrain
the kinematics of squark and gluino production.

The fit proposed in this paper is reasonably model independent
compared to the previous fits using the entire $A_T$ distribution
without $m_{ll}$ cuts.  It is amazing to see that the distribution
keeps the information on the cascade decay kinematics. (Compare
Fig.~7 and 10). The analysis
can be extended to all cascade decays involving leptons, such as the
gauge mediated scenario with NLSP slepton.\cite{DTW, GM} 
The determination of the $A_T$ peak position is not
disturbed even in the case where several sleptons contribute to signal
lepton pairs.

Note that model independent constraints on weakly interacting
sparticle masses may be used to directly constrain the relic mass
density of LSPs in our Universe. The density of such Big Bang relics
is roughly proportional to the inverse of the pair annihilation cross
section of the lightest neutralino. In the MSUGRA model,
$1/\sigma\sim$ $m^4_{\tl}/m^2_{\tchi^0_1}$ in the bino dominant
limit.\cite{DM}
If the overall sparticle scale is constrained within 10\%, an
upper bound on the mass density could be derived within 20\%. The
improved determination of SUSY parameters at the LHC combined with
improved astronomical observations might significantly constrain the
remaining MSSM parameters.

In this paper, we did not perform any MC simulation for Tevatron
experiments. There the cleanest discovery process is the 3 leptons and
missing $\esla_T$ channel of $\tchi^+_1\tchi^0_2$ production and
decay. It is possible to perform a parallel analysis to the one
presented in this paper. However if $m^{\rm max}_{ll}$ is small
(which is likely due to the lower bound on $m_{\tilde l}$ of nearly 100 GeV),
the number of events that satisfy $12$ GeV $<m_{ll}<m^{\rm max}_{ll}/2$
would be small, where the lower $m_{ll}$ cut is needed to avoid $\gamma^*$
and $\Upsilon$ backgrounds.

The branching ratio of the mode $\tchi^0_2\rightarrow\tilde{l}
l$ could be small if other modes such us $\tchi^0_2\rightarrow
Z, h...$ dominate. The decay $\tchi^0_2\rightarrow \tilde{\tau} \tau$
may be only two body decay channel accessible in MSUGRA model due to
$\tilde{\tau}$ mixing.  The analysis would be substantially more
difficult for this case, as $\tau$ decays further into a jet or a
lepton.\cite{tau} Selecting two tau leptons which go roughly into the same
direction (small $\Delta R$) should effectively work as an $m_{\tau\tau}$
cut in our analysis. However, the $A_T$ distribution of tau jet would
be substantially smeared by the tau decay.

When all two body decay modes are closed, the decay
$\tchi^0_2\rightarrow$ $\tilde{\chi}^0_1 ll$ often has a sizable
branching ratio. The precision study of the three body decay
distribution has been discussed in \cite{NY}.  The $m_{ll}$
distribution and the $A_{T}$ distribution in the small $m_{ll}$ region
would give us information on neutralino mixing and on
$m_{\tl_{L(R)}}$.

It would be interesting to check if our analysis can be extended to
other cascade decays involving photons or jets \cite{GM}. Note that in
the gauge mediated model with $\tchi^0_1$ NLSP, the decay chain
$\tchi^0_2\rightarrow\tchi^0_1 ll$ may be associated with a photon
from $\tchi^0_1\rightarrow \tilde{G} \gamma$ \cite{DTW}. Cascade
decays involving a jet and a lepton or two jets may also be used for
an asymmetry analysis, but selecting the proper combination of jets
would be challenging.

\section*{Acknowledgments}
We thank H. Baer and M. Drees for discussions. We also thank M. Drees for 
careful reading of the manuscript. M. N. with to  thank to ITP, Santa Barbara 
for its support during part of this work (NSF Grant No. PHY94-07194). 

\end{document}